\documentclass[onecolumn]{revtex4}

\usepackage{graphicx}
\usepackage{dcolumn}
\usepackage{bm}
\usepackage{color}
\usepackage{amssymb,amsmath}

\input epsf

\begin{document}

\title{\Large{Dilaton Dark Energy Model in $f(R)$, $f(T)$ and Ho$\check{\text r}$ava-Lifshitz Gravities}}

\author{\bf Piyali Bagchi
Khatua$^1$\footnote{piyali.bagchi@yahoo.co.in}, Shuvendu
Chakraborty$^2$\footnote{shuvendu.chakraborty@gmail.com} and Ujjal
Debnath$^3$\footnote{ujjaldebnath@yahoo.com}}
\affiliation{$^1$Department of CSE, Netaji Subhas Engineering
College, Garia, Kolkata-700 152, India.\\
$^2$Department of Mathematics,Seacom Engineering College, Howrah,
711 302, India.\\
$^3${Department of Mathematics, Bengal Engineering and Science
University, Shibpur, Howrah-711 103, India.} }

\date{\today}

\begin{abstract}
In this work, we have considered dilaton dark energy model in
Weyl-scaled induced gravitational theory in presence of barotropic
fluid. It is to be noted that the dilaton field behaves as a
quintessence. Here we have discussed the role of dilaton dark
energy in modified gravity theories namely, $f(R), f(T)$ and
Ho$\check{\text r}$ava-Lifshitz gravities and analyzed the
behaviour of the dilaton field and the corresponding potential in
respect to these modified gravity theories instead of Einstein's
gravity. In $f(R)$ and $f(T)$ gravities, we have considered some
particular forms of $f(R)$ and $f(T)$ and we have shown that the
potentials always increase with the dilaton fields. But in
Ho$\check{\text r}$ava-Lifshitz gravity, it has been seen that the
potential always decreases as dilation field increases.\\
\end{abstract}

\maketitle
\section{\normalsize\bf{Introduction}}

Recent observations indicate that our universe is currently
expanding with an acceleration [1]. The main responsible candidate
for this acceleration is dark energy (DE). There are several types
of exotic type of dark energy with negative pressure depending on
their equation of state (EOS). When EOS $-1<w<-1/3$, it is called
quintessence and when $w<-1$, it is phantom. There are some other
dark energies model which can cross the phantom divide $w=-1$ both
sides are called quintom. Recently such acceleration is understood
by imposing a concept of modification of gravity for an
alternative candidate of dark energy [2 - 40]. This concept
provides very natural gravitational alternative for exotic
matter. This type of gravity is predicted by string/M-theory. The
explanation of the phantom or non-phantom or quintom phase of the
universe can be described by this gravity without introducing
negative kinetic term of dark energies. After proposing dark
energy models to explain cosmic acceleration it was realized that
even non linear terms of Ricci curvature $R^{-n} (n > 0)$ can be
used as an alternative of DE [2 - 15]. Several authors considered
different form of $f(R)$ which exhibit late time acceleration for
small curvature and inflation for large curvature. The modified
gravity theory can be constructed by adding geometrical
correction terms to the usual Einstein-Hilbert Lagrangian $L$
considering as a function of scalar curvature $L=f(R)$. $f(R)$
model has the importance that it
satisfy both the cosmological as well as local gravity constraints.\\

Another approach can be explored using Weitzenb\"{o}ck connection
having no curvature but torsion which is formed from products of
first derivatives of the tetrad with no second derivatives in the
torsion tensor. It is extremely relate to general relativity only
exception in boundary terms [16 - 22]. In this modified gravity
approach $f(T)$ torsion will be responsible candidate of the
observed acceleration of the universe without restoring the DE. An
advantage of the generalized $f(T)$ theory than $f(R)$ theory is
its field equations of second order compare to forth
ordered equations of $f(R)$ theory.\\

Recently Ho$\check{\text r}$ava [23] proposed a new theory of
gravity. It is renormalizable with higher spatial derivatives in
four dimensions which reduces to Einstein's gravity with non
vanishing cosmological constant in IR but with improved UV
behaviours. It is similar to a scalar field theory of Lifshitz
[24] in which the time dimension has weight 3 if a space
dimension has weight 1and this theory is called Ho$\check{\text
r}$ava-Lifshitz gravity. Ho$\check{\text r}$ava-Lifshitz gravity
has been studied and extended in detail and applied as a
cosmological
framework of the universe [25 - 40].\\

One can construct a dilaton dark energy model as non-minimal
quintessence based on Weyl-scaled induced gravitational theory to
find an important thing that when dilaton field is not
gravitational clustered at small scales, the effect of dilation
can not change the evolutionary law of structure formation [41,
42]. Quintessence dark energy models have been widely studied and
here we consider dilaton dark energy model in Weyl-scaled induced
gravitational theory as a quintessence [41 - 56] in $f(R), f(T)$
and Ho$\check{\text r}$ava-Lifshitz gravities and analyze the
behaviour of the potential and field in respect to these modified dark energy models.\\

\section{\bf{Basic Equations in Dilaton Dark Energy Model}}

The action of the Weyl-scaled induced gravitational theory is
given by [42,45]
\begin{equation}
 S =\int d^{4}x\sqrt{-g}\left[\frac{1}{2}R(g_{\mu\nu})-\frac{1}{2}g^{\mu\nu}
 \partial_{\mu}\sigma\partial_{\nu}\sigma-V(\sigma)+\frac{1}{2}g^{\mu\nu}
 e^{-\alpha\sigma}\partial_{\mu}\phi\partial_{\nu}\phi-e^{-2\alpha\sigma}V(\phi)\right]
\end{equation}

where $\alpha=\sqrt{\frac{\kappa^2}{2\varpi+3}}$ with $\varpi$
being an important parameter in Weyl-scaled induced gravitational
theory. Here $\sigma$ is the dilation field and $V(\sigma)$ is
the scalar potential of the dilation field. \\

By varying action (1) and considering dilaton field as the
candidate of DE, the field equations and the conservation
equations in Friedmann-Robertson-Walker model, become
($\kappa^{2}\equiv 8\pi G$):
\begin{equation}
H^2+\frac{k}{a^{2}} = \frac{\kappa^2}{3}(\rho_{m}+\rho_{\sigma})
\end{equation}
\begin{equation}
\dot{H}-\frac{k}{a^{2}}=-\frac{\kappa^2}{2}(\rho_{m}+p_{m}+\rho_{\sigma}+p_\sigma)
\end{equation}
\begin{equation}
\dot{\rho}_{m}+3H(\rho_{m}+p_{m})=\frac{1}{2}\alpha\dot{\sigma}(\rho_{m}+3p_{m})
\end{equation}
 and
\begin{equation}
\dot{\rho}_{\sigma}+3H\dot{\sigma}^{2}=\frac{1}{2}\alpha
e^{-\alpha\sigma}(\rho_{m}-3p_{m})
\end{equation}

where $H=\frac{\dot{a}}{a}$ is the Hubble parameter, $\rho_{m}$ is
dark matter energy density, $\rho_{\sigma}$ is dilaton dark energy
density and radiation is neglected. The effective energy density
and pressure of dilaton dark energy are given by,
\begin{equation}
\rho_{\sigma}=\frac{1}{2}\dot{\sigma}^2+V(\sigma)
\end{equation}
and
\begin{equation}
p_{\sigma}=\frac{1}{2}\dot{\sigma}^2-V(\sigma)
\end{equation}
 For matter, $p_{m}=0$ we have from equation (4)
$\rho_{m}=\rho_{m0} a^{-3}e^{\frac{\alpha\sigma}{2}}$ as the
matter density. For perfect fluid with barotropic equation of
state $p_{m}=w_{m}\rho_{m}$, we get
the density of the fluid as $\rho_{m}=\rho_{m0} a^{-3(1+w_{m})}e^{\frac{\alpha(1+3w_{m})\sigma}{2}}$.\\

In next three sections we analyze the effect of dilaton dark
energy in three modified gravity models mentioned above.

\section{\bf{Dilaton Dark Energy Model in $f(R)$ Gravity}}

In the four-dimensional flat space-time, the action of $f(R)$
gravity with matter can be written as [5],
\begin{equation}
I=\int
d^4x\sqrt{-g}\left[\frac{f(R)}{2\kappa^2}+{\cal{L}}_m\right]
\end{equation}
where the usual Einstein-Hilbert action is generalized by
replacing $R$ with $f(R)$, which is an analytic function of $R$
and $g$ is the determinant of the metric tensor $g_{\mu\nu}$ and
${\cal{L}}_m$ is the matter Lagrangian. In this flat space-time
($k=0$) the Ricci scalar is given by $R=6\dot{H}+12H^2$.\\

From the variation of action (8), we obtain the field equations of
this modified gravity is given by,

\begin{equation}
f'(R) R_{\mu\nu}- \frac{1}{2}g_{\mu\nu}f(R)+ g_{\mu\nu}\Box
f'(R)-\nabla_{\mu}\nabla_{\nu} f'(R) = \kappa^2 T^{m}_{\mu \nu}
\end{equation}

where $T^{m}_{\mu \nu}$ are the energy momentum tensor components
the matter field and prime denotes differentiation with respect to
the scalar curvature $R$.
  The gravitational field equations in flat ($k=0$) space-time take the
form
\begin{equation}
H^2=\frac{\kappa^2}{3f'(R)}(\rho_{m}+\rho_{c})
\end{equation}
and
\begin{equation}
\dot{H}=-\frac{\kappa^2}{2f'(R)}(\rho_{m}+p_{m}+\rho_c+p_c)
\end{equation}
where $\rho_m$ being the energy density and $p_m$ is the pressure
of all ordinary matter. $\rho_c$ and $p_c$ can be regarded as the
energy density and pressure generated due to the difference of
$f(R)$ gravity and general relativity, given by [4]
\begin{equation}
\rho_{c}=\frac{1}{\kappa^2}\left[\frac{1}{2}\left(-f(R)+Rf'(R)\right)-3H\dot{R}f''(R)\right]
\end{equation}
and
\begin{equation}
p_{c}=\frac{1}{\kappa^2}\left[\frac{1}{2}\left(f(R)-Rf'(R)\right)+\left(2H\dot{R}+\ddot{R}\right)f''(R)+\dot{R}^2f'''(R)\right]
\end{equation}
Now from (2), (3), (6), (7), (10) and (11) we get,
\begin{equation}
\rho_{m}+\rho_{c}=f'(R)\left(\rho_{m}+\frac{1}{2}\dot{\sigma}^2+V(\sigma)\right)
\end{equation}
and
\begin{equation}
\rho_{m}+p_{m}+\rho_{c}+p_c=f'(R)\left(\rho_{m}+p_{m}+\dot{\sigma}^2\right)
\end{equation}

Now consider the model where $f(R)$ is defined as,
\begin{equation}
f(R)=R+\xi R^\mu+\eta R^{-\nu}
\end{equation}
Taking the higher derivatives of the above equation w.r.t. $R$ we
get,
$$f'(R)=1+\xi \mu R^{\mu-1}-\eta \nu R^{-\nu-1}$$
$$f''(R)=\xi \mu (\mu-1) R^{\mu-2}+\eta \nu (\nu+1) R^{-\nu-2}$$
and $$f'''(R)=\xi \mu (\mu-1) (\mu-2) R^{\mu-3}-\eta \nu (\nu+1)
(\nu+2) R^{-\nu-3}$$ Now we take the Hubble parameter $H$ of the
form, $H=a_0a^{n}$
where $n>0$. Thus we have, $\dot{H}=na_0^2a^{2n}$.\\

Now using the equations (12) - (16) we get,
\begin{eqnarray*}
V(\sigma)=-(2a^{2n}a_0^2(\eta(1+\nu)(6+n(3+\nu(-5+n+2n\nu)))-6^{\mu+\nu}
(-a^{2n}a_0^2(2+n))^{\mu+\nu}(-1+\mu)(6+n(3+\mu(5+n
\end{eqnarray*}
\begin{eqnarray*}
(-1+2\mu))))\xi)+e^{\frac{\alpha\sigma}{2}}\kappa^2(\eta\nu-\nu(2^{1+\mu+\nu}3^{\mu+\nu}-6^{\mu+\nu})
(-a^{2n}a_0^2(2+n))^{\mu+\nu}\mu\xi)\rho_0)
\end{eqnarray*}
\begin{equation}
/(2\kappa^2(6a^{2n}a_0^2(-a^{2n}a_0^2(2+n))^{\nu}(2^{1+\nu}3^\nu+6^\nu
n)+\eta\nu-6^{\mu+\nu} (-a^{2n}a_0^2(2+n))^{\mu+\nu}\mu\xi))
\end{equation}
and
\begin{eqnarray*}
\sigma=\int\sqrt{\frac{1}{\kappa^2}(2a^{2n}a_0^2
n(\eta\nu(1+\nu)(1+n+2n\nu)-6^{\mu+\nu}(-a^{2n}a_0^2(2+n)^{\mu\nu}(-1+\mu)
\mu(-1+n(-1+2\mu))\xi)-a^{-3}e^\frac{\alpha\sigma}{2}\kappa^2}
\end{eqnarray*}
\begin{equation}
\overline{(\eta\nu-6^{\mu+\nu}(-a^{2n}a_0^2(2+n))^{\mu+\nu}\mu\xi)\rho_0))/6a^{2n}a_0^2
(-a^{2n}a_0^2(2+n))^\nu(2^{1+\nu}3^\nu +6^\nu
n)+\eta\nu-6^{\mu+\nu}(-a^{2n}a_0^2(2+n))^{\mu+\nu}\mu\xi}~dt
\end{equation}

\begin{figure}[h!]
\includegraphics[height=3.0in]{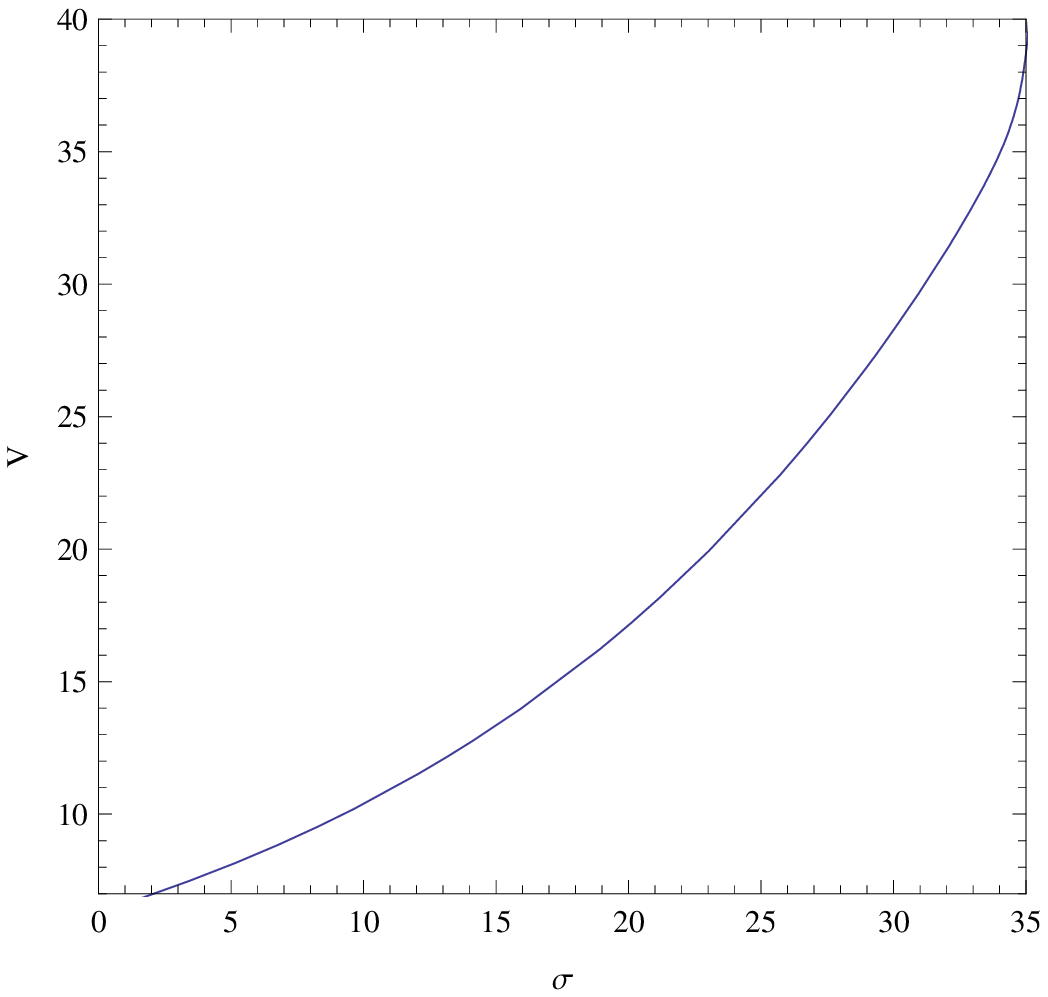}~\\
~~~Fig.1~~~\\
\vspace{2mm}Fig.1 shows the variation of $V$ against $\sigma$ for
$a_0=1, n=1.5$.

\vspace{6mm}

\end{figure}

From above, we have seen that the expressions of $\sigma$ and $V$
are very complicated. So $V$ cannot be expressed explicitly in
terms of $\sigma$. Fig.1 shows the the variation of $V$ against
$\sigma$ for $a_0=1, n=1.5$ in $f(R)$ gravity model. From the
figure, we have seen that the potential $V$ always increases
with the dilaton field $\sigma$.\\

\section{\bf{Dilaton Dark Energy Model in $f(T)$ Gravity}}
The action for the $f(T)$ gravity is given by [19],
\begin{equation}
S=\frac{1}{2\kappa^2}\int dx^4
\left[\sqrt{-g}f(T)+{\cal{L}}_m\right]
\end{equation}
where $T$ is the torsion scalar, $f(T)$ is general differentiable
function of the torsion. Here the torsion scalar $T$ is defined as
[9],
\begin{equation}
T=S^{\mu\nu}_{\rho}T^{\rho}_{\mu\nu}
\end{equation}
where,
\begin{equation}
S^{\mu\nu}_{\rho}=\frac{1}{2}(\delta^{\mu}_{\rho}{T^{\theta\nu}}_{\theta}-
\delta^{\nu}_{\rho}{T^{\theta\mu}}_{\theta})-\frac{1}{4}({T^{\mu\nu}}_{\rho}-{T^{\nu\mu}}_{\rho}-T_{\rho}^{\mu\nu})
\end{equation}
and
\begin{equation}
T^{\lambda}_{\mu\nu}=e^{\lambda}_{i}(\partial_\mu
e^{i}_{\nu}-\partial_\nu e^{i}_{\mu})
\end{equation}
where $e=\sqrt{-g}$. We now assume a flat ($k=0$) homogeneous and
isotropic FRW universe and for this model,
\begin{equation}
e^i_\mu=diag(1,a(t),a(t),a(t))  \qquad\qquad and \qquad\qquad
T=-6H^2
\end{equation}
 The modified Fridmann equations can be written as,
\begin{equation}
12H^2 f'(T)+f(T)=2\kappa^2\rho_{m}
\end{equation}
and
\begin{equation}
48H^2\dot{H}f''(T)-(12H^{2}+4\dot{H})f'(T)-f(T)=2\kappa^2(\rho_{m}+p_{m})
\end{equation}
where prime denotes the derivatives w.r.t. $T$.\\

Now putting $T=-6H^2$, the above set of equations becomes,
\begin{equation}
-2Tf'(T)+f(T)=2\kappa^2\rho_{m}
\end{equation}
and
\begin{equation}
-8T\dot{H}f''(T)+(2T-4\dot{H})f'(T)-f(T)=2\kappa^2(\rho_{m}+p_{m})
\end{equation}
As the equation (26) and (27) will become usual Einstein's
equations for a special case when $f(T)=T$ so the equations can be
rewritten as
\begin{equation}
H^2=\frac{\kappa^2}{3}(\rho_{m}+\rho_{T})
\end{equation}
and
\begin{equation}
\dot{H}=-\frac{\kappa^2}{2}(\rho_{m}+p_{m}+\rho_{T}+p_{T})
\end{equation}
where $\rho_{T}$ and $p_{T}$ be the torsion contributions to the
energy density and pressure in $f(T)$ gravity.\\

From the above equations we get,
\begin{equation}
\rho_{T}=\frac{1}{2\kappa^2}(2Tf'(T)-f(T)+6H^2),~~p_T=-\frac{1}{2\kappa^2}(-8
\dot{H}Tf''(T)+(2T-4\dot{H})f'(T)-f(T)+4\dot{H}+6H^2)
\end{equation}

Now consider a particular modified $f(T)$ gravity model where
$f(T)$ is defined as,
\begin{equation}
f(T)=\beta T+\gamma T^{m}
\end{equation}
Taking the higher derivatives of the above equation w.r.t. $T$ we
get,
$$f'(T)=\beta +m\gamma T^{m-1}$$
$$f''(T)=m(m-1)\gamma T^{m-2}$$
and $$f'''(T)=m(m-1)(m-2)\gamma T^{m-3}$$
 Now we take $H$ of the form,
$H=a_0a^{-n}$ where $a_0$ is a constant and $n>0$.\\

Now from (2), (3), (30), (31) and using the above derivatives we
get,
\begin{equation}
V(\sigma)=\frac{a^{-2n}}{6a_0^2\kappa^2}(6a_0^4(3+(-3+n)\beta)-6^m
a^{4n}(-a^{-2n}a_0^2)^m
m\gamma+a^{2n}a_0^2(6\beta-6^m(-a^{2n}a_0^2)^m(-1+2m)(-3+m
n)\gamma))
\end{equation}
and
\begin{equation}
\sigma=\int{\sqrt{\frac{(-6-6a^{-2n}a_0^2n)\beta+\frac{1}{a_0^2}
(-a^{-2n}a_0^2)^m m(6^m a^{2n}-a_0^2(6^m-2^{1+m}3^m
m)n)\gamma}{3\kappa^2}}}dt
\end{equation}

\begin{figure}[h!]
\includegraphics[height=3.5in]{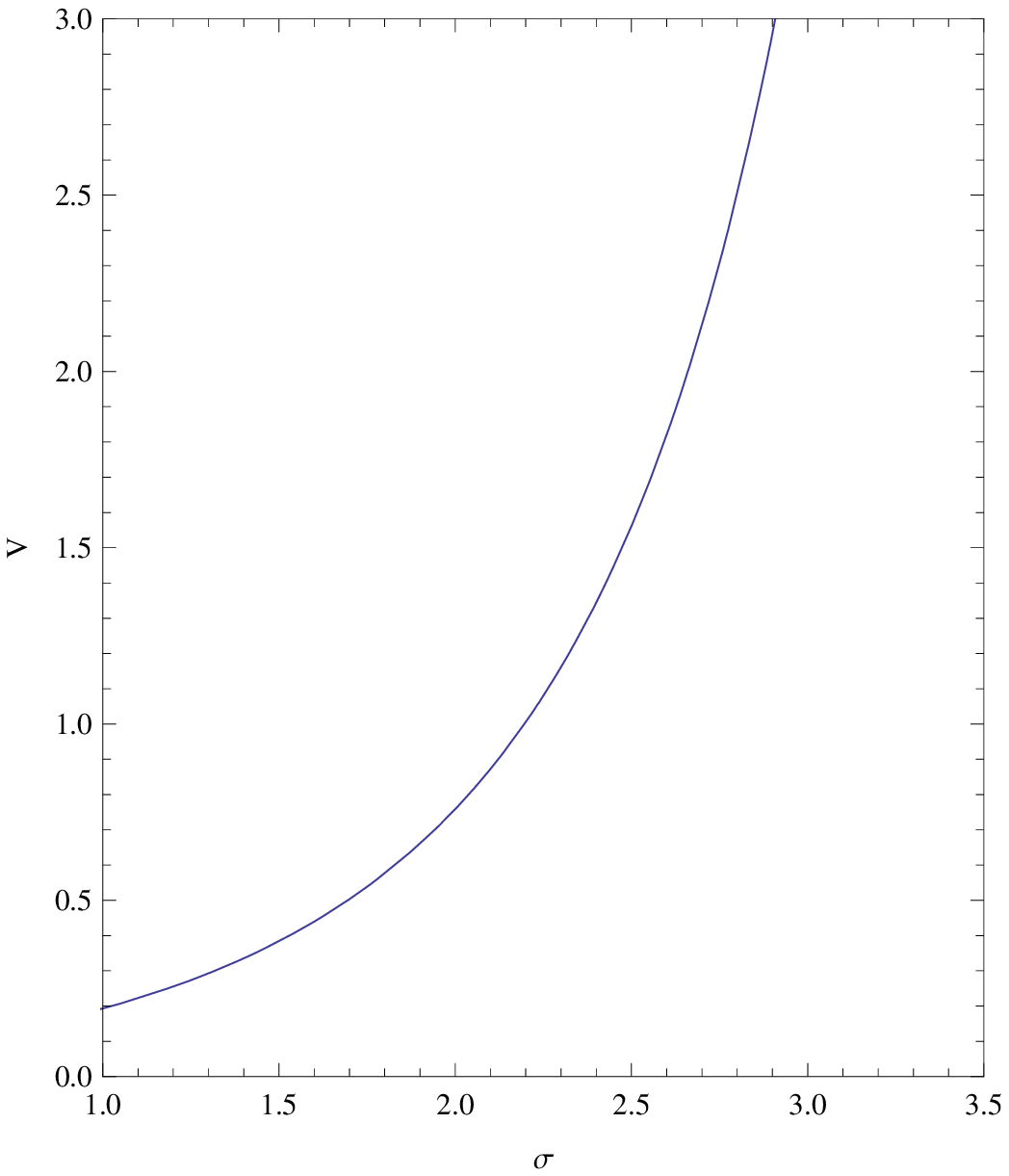}~\\
~~~Fig.2~~~\\
 \vspace{2mm}Fig.2 represents the variation of $V$ against $\sigma$
for $a_0=1, n=1.5$.

\vspace{6mm}

\end{figure}

From above, we have seen that the expressions of $\sigma$ and $V$
are very complicated. So $V$ cannot be expressed explicitly in
terms of $\sigma$. Fig.2 represents the variation of $V$ against
$\sigma$ for $a_0=1, n=1.5$ in $f(T)$ gravity model. From the
figure, we have seen that $V$ always increases with $\sigma$.\\

\section{\bf{Dilaton Dark Energy Model in Ho$\check{\text r}$ava-Lifshitz Gravity}}

In the (3+1) dimensional Arnowitt-Deser-Misner formalism the full
metric is written as [39],

\begin{equation}
ds^{2}=-N^{2}dt^{2} + g_{ij}(dx^{i} + N^{i}dt)(dx^{j} + N^{j}dt)
\end{equation}

Under the detailed balance condition the full action condition of
Ho$\check{\text r}$ava-Lifshitz gravity is given by,

\begin{eqnarray*}
S=\int dt d^{3}x \sqrt{g}N
\left[\frac{2}{\kappa_{1}^{2}}(K_{ij}K^{ij} - \lambda K^{2}) +
\frac{\kappa_{1}^{2}}{2 \omega^{4}}C_{ij}C^{ij}-
\frac{\kappa_{1}^{2}\mu \epsilon^{ijk}}{2 \omega^{2} \sqrt{g}}
R_{il}\nabla_{j}R^{l}_{k}\right.\end{eqnarray*}
\begin{equation}
\left.+ \frac{\kappa_{1}^{2}\mu^{2}}{8}R_{ij}R^{ij}+
\frac{\kappa_{1}^{2}\mu^{2}}{8(3 \lambda - 1)}\left(\frac{1 -4
\lambda}{4}R^{2} + \Lambda R - 3 \Lambda^{2} \right)\right]
\end{equation}

where

\begin{equation}
K_{ij}=\frac{1}{2N}(\dot{g}_{ij} - \nabla_{i}N_{j}-
\nabla_{j}N_{i})
\end{equation}

is the extrinsic curvature and

\begin{equation}
C^{ij}=\frac{\epsilon^{ikl}}{\sqrt{g}}\nabla_{k}(R^{j}_{i} -
\frac{1}{4}R \delta^{j}_{l})
\end{equation}

is known as Cotton tensor and the covariant derivatives are
defined with respect to the spatial metric $g_{ij}$.
$\epsilon^{ijk}$ is the totally antisymmetric unit tensor,
$\lambda$ is a dimensionless coupling constant and the variable
$\kappa_{1}$ , $\omega$ and $\mu$ are constants with mass
dimensions $-1,~ 0,~ 1$ respectively.\\

Now, in order to focus on cosmological frameworks, we impose the
so called projectability condition and use a FRW metric we get,

\begin{equation}
N=1,    g_{ij}=a^{2}(t)\gamma_{ij},      N^{i}=0
\end{equation}

with

\begin{equation}
\gamma_{ij} dx^{i} dx^{j}= \frac{dr^{2}}{1 - kr^{2}}+
r^{2}d\Omega^{2}_{2},
\end{equation}
where $k=0, -1, +1$ corresponding to flat, open and closed
respectively. By varying $N$  and $g_{ij}$, we obtain the
non-vanishing equations of motions:

\begin{equation}
H^{2}=\frac{\kappa_{1}^{2}}{6(3\lambda -1)} ~ \rho_{m} +
\frac{\kappa_{1}^{2}}{6(3\lambda -1)} \left[\frac{3 \kappa^{2}
\mu^{2} k^{2}}{8(3\lambda -1)a^{4}} + \frac{3 \kappa^{2} \mu^{2}
\mu^{2} \Lambda^{2}}{8(3\lambda -1)}\right] - \frac{ \kappa^{4}
\mu^{2}k \Lambda }{8(3\lambda -1)^{2}a^{2}}
\end{equation}
and
\begin{equation}
\dot{H} + \frac{3}{2}H^{2}= - \frac{\kappa_{1}^{2}}{4(3\lambda
-1)}~p_{m} - \frac{\kappa_{1}^{2}}{4(3\lambda -1)} \left[\frac{
\kappa^{2} \mu^{2} k^{2}}{8(3\lambda -1)a^{4}}- \frac{3 \kappa^{2}
\mu^{2} \Lambda^{2}}{8(3\lambda -1)}\right] - \frac{ \kappa^{4}
\mu^{2}k \Lambda }{16(3\lambda -1)^{2}a^{2}}
\end{equation}

The term proportional to $a^{-4}$ is the usual ``dark radiation",
present in Ho$\check{\text r}$ava-Lifshitz cosmology while the
constant term is just the explicit cosmological constant. For
$k=0$, there is no contribution from the higher order derivative
terms in the action. However for $k\ne 0$, their higher derivative
terms are significant for small volume i.e., for small $a$ and
become insignificant for large $a$, where it agrees with general
relativity. As a last step, requiring these expressions to
coincide the standard Friedmann equations, in units where $c=1$,

\begin{equation}
G_{c}=\frac{\kappa_{1}^{2}}{16 \pi(3\lambda -1)}
\end{equation}

\begin{equation}
\frac{ \kappa_{1}^{4} \mu^{2} \Lambda }{8(3\lambda -1)^{2}}=1
\end{equation}

where $G_{c}$ is the ``cosmological" Newton's constant. We mention
that in theories with Lorentz invariance breaking (such is
Ho$\check{\text r}$ava-Lifshitz one) the ``gravitational" Newton's
constant $G$, that is the one that is present in the gravitational
action, does not coincide with $G_{c}$, that is the one that is
present in Friedmann equations, where

\begin{equation}
G=\frac{\kappa_{1}^{2}}{32 \pi}
\end{equation}

as it can be straightforwardly read from the action. In the IR
$(\lambda=1)$ where Lorentz invariance is restored, $G_{c}=G$.
Using the above identifications, we can re-write the Friedmann
equations as,

\begin{equation}
H^{2} + \frac{k}{a^{2}}=\frac{l^2}{3}\rho_m + \frac{k^{2}}{2
\Lambda a^{4}} + \frac{\Lambda}{2}
\end{equation}
and
\begin{equation}
\dot{H} - \frac{k}{a^{2}}= -\frac{l^2}{2}(\rho_{m}+p_m) -
\frac{k^{2}}{\Lambda a^{4}}
\end{equation}

where $l^2=8\pi G_{c}~$. Again the usual Einstein's field
equations are
\begin{equation}
H^{2}+ \frac{k}{a^{2}}=\frac{\kappa^2}{3}(\rho_{m}+\rho_h)
\end{equation}
and
\begin{equation}
\dot{H}-
\frac{k}{a^{2}}=-\frac{\kappa^2}{2}(p_{m}+p_h+\rho_{m}+\rho_{h})
\end{equation}
Here, $\rho_{h}$ and $p_h$ are the contribution of energy density
and pressure by Ho$\check{\text r}$ava-Lifshitz gravity. Now from
(45) and (47) we get,
\begin{equation}
\rho_h=\left(\frac{l^{2}}{\kappa^{2}}-1
\right)\rho_{m}+\frac{3}{2\kappa^{2}}\left(\frac{k^{2}}{\Lambda
a^{4}}+\Lambda \right)
\end{equation}

Equating (46) and (48) we get,
\begin{equation}
p_h+\rho_h=\left(\frac{l^{2}}{\kappa^{2}}-1
\right)(\rho_{m}+p_{m})+\frac{k^{2}}{\kappa^{2}\Lambda a^{4}}
\end{equation}

Thus $\rho_{\sigma}=\rho_{h}$ and $p_{\sigma}=p_h$ give,

\begin{equation}
\sigma=\int\sqrt{\left(\frac{l^{2}}{\kappa^{2}}-1
\right)(\rho_{m}+p_{m})+\frac{k^{2}}{\kappa^{2}\Lambda a^{4}}}~dt
\end{equation}
and
\begin{equation}
V(\sigma)=\frac{1}{2}\left(\frac{l^{2}}{\kappa^{2}}-1
\right)(\rho_{m}-p_{m})+\frac{k^{2}}{\kappa^{2}\Lambda
a^{4}}+\frac{3\Lambda}{2\kappa^{2}}
\end{equation}

\begin{figure}
\includegraphics[height=3.0in]{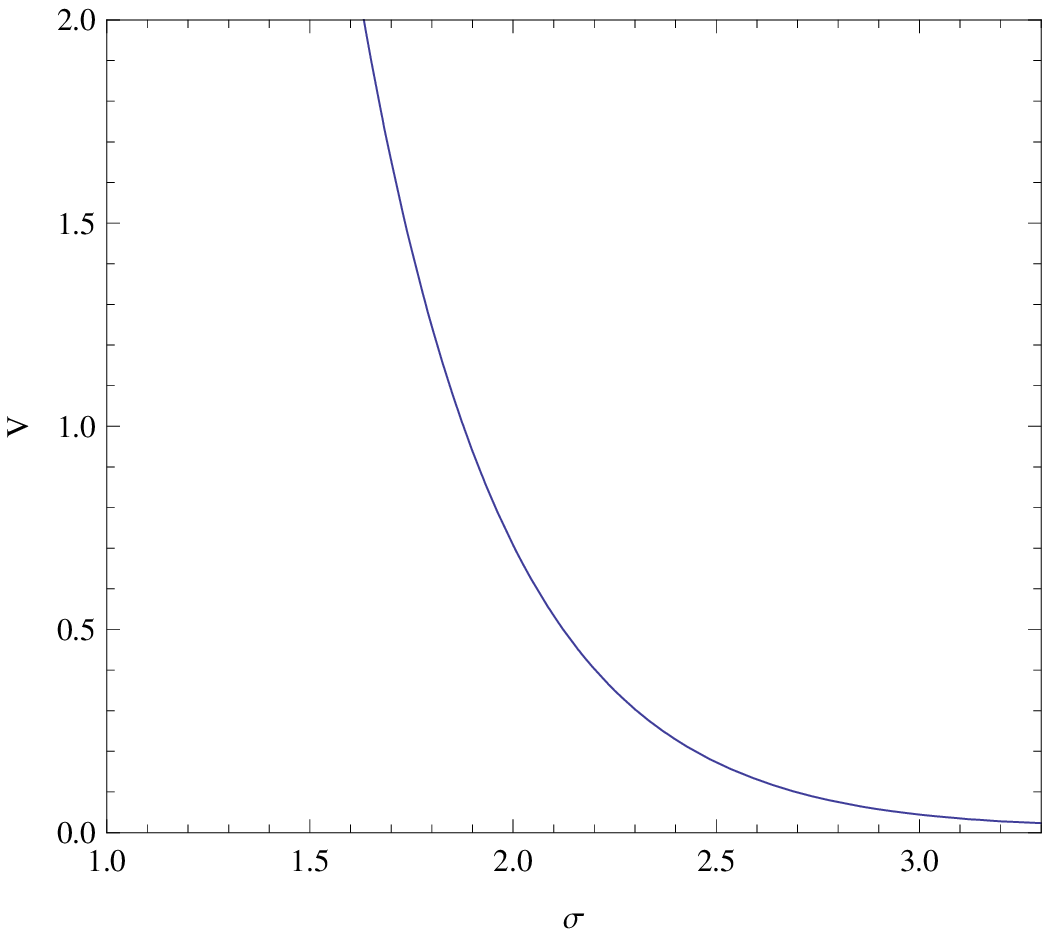}~\\
~~Fig.3~~~\\
\vspace{2mm}Fig.3 represents the variation of $V$ against $\sigma$
for $\rho_0=1, k=1$.

\vspace{6mm}

\end{figure}

Fig.3 ~represents the variation of $V$ against $\sigma$ for
$\rho_0=1, k=1$ in Ho$\check{\text r}$ava-Lifshitz gravity model.
From the figure, we have seen that $V$ always increases with
$\sigma$.\\

\section{\normalsize\bf{Discussions}}

Our aim in this work is to construct a cosmological model by using
modified gravity theories of some particular forms as the
contribution of a effective dark energy. Also an additional dark
fluid has been included to play an important cosmological role. In
fact we consider two accelerated epoch of the universe described
by a model where the mysterious energy generated from two separate
source. So in our model, the acceleration of the universe become
faster than the normal dark energy effect or modified gravity
effect separately and give rise to an super acceleration phase. In
this work, we have considered dilaton dark energy model in
Weyl-scaled induced gravitational theory in presence of barotropic
fluid. It is to be noted that the dilaton field behaves as a
quintessence. Here we have discussed the role of dilaton dark
energy in modified gravity theories namely, $f(R), f(T)$ and
Ho$\check{\text r}$ava-Lifshitz gravities and graphically analyzed
the behaviour of the dilaton field and the corresponding potential
in respect to these modified gravity theories instead of
Einstein's gravity. In $f(R)$ and $f(T)$ gravities, we have
considered some particular forms of $f(R)$ and $f(T)$ and we have
shown that the potentials always increase with the dilaton fields.
But in Ho$\check{\text r}$ava-Lifshitz gravity, it has been seen
that the
potential always decreases as dilation field increases.\\\\

{\bf Acknowledgement:}\\

The authors are thankful to IUCAA, Pune, India for warm
hospitality where part of the work was carried out.\\\\

{\bf References:}\\

[1] A. G. Riess et al,  \textit{Astron. J.} \textbf{116} 1009 (1998).\\\

[2] S. Capozziello et al, \textit{IJMPD} \textbf{12} 1969 (2003).\\\

[3] S. M. Carroll et al,  \textit{Phys. Rev. D} \textbf{70} 043528 (2004).\\\

[4] H. M. Sadjadi, \textit{Phys. Rev. D} \textbf{77} 103501
(2008); \textit{Phys. Rev. D} \textbf{76} 104024 (2007).\\\

[5] S. Nojiri and S. D. Odintsov, \textit{arXiv:}
\textbf{0807.0685v1} [hep-th] (2008); \textit{Phys. Rev. D}
\textbf{78} 046006 (2008);  \textit{Phys. Rev. D} \textbf{77}
026007 (2008); \textit{Int. J. Geom. Meth. Mod. Phys.} \textbf{ 4} 115 (2007).\\\

[6] V. Faraoni, \textit{Phys. Rev. D} \textbf{75} 067302 (2007).\\\

[7] M. Akbar and R. G. Cai,  \textit{Phys. Lett. B } \textbf{635} 7 (2006).\\\

[8] G. Cognola et al, \textit{Phys. Rev. D} \textbf{79} 044001 (2009).\\\

[9] E. Elizalde et al, \textit{Phys. Rev. D} \textbf{80} 044030 (2009).\\\

[10] A. D. Felice and S. Tsujikawa, \textit{Living Rev. Rel.} \textbf{13} 3 (2010).\\\

[11] K. Bamba and C. Q. Geng, \textit{Phys. Lett. B} \textbf{679} 282 (2009).\\\

[12] S. K. Srivastava, \textit{Phys. Lett. B} \textbf{648} 119
(2007); \textit{Phys. Lett. B} \textbf{643} 1 (2006);
\textit{Int. J. Theor. Phys.} \textbf{47} 1966 (2008);
\textit{arXiv:} \textbf{0511167v5} [astro-ph] (2006);
\textit{arXiv:} \textbf{0809.1950v1} [gr-qc] (2008);
 \textit{Phys. Rev. A} \textbf{78} 062322 (2008).\\\

[13] S. Das et al, \textit{Class. Quant. Grav.} \textbf{23} 4159 (2006).\\\

[14] D. N. Vollick, \textit{Phys. Rev. D} \textbf{76} 124001 (2007).\\\

[15] Y. Bisabr, \textit{Phys. Lett. B} \textbf{690} 456 (2010).\\\

[16] R. Ferraro and F. Fiorini,  \textit{Phys. Rev. D} \textbf{78} 124019 (2008).\\\

[17] R. Ferraro and F. Fiorini,  \textit{Phys. Rev. D} \textbf{75} 084031 (2007).\\\

[18] E. V. Linder, \textit {Phys. Rev. D} \textbf{ 81} 127301 (2010).\\\

[19] K. K. Yerzhanov et al, \textit{arXiv:} \textbf{1006.3879v1} [gr-qc] (2010).\\\

[20] S. H. Chen, \textit{Phys. Rev. D} \textbf{83} 023508 (2011).\\\

[21] G. R. Bengochea, \textit{Phys. Rev. D} \textbf{79} 124019 (2009).\\\

[22] R. Myrzakulov, \textit{arXiv:} \textbf{1006.1120v1} [gr-qc] (2010).\\\

[23] P. Ho$\check{\text r}$ava,  \textit{JHEP} \textbf{0903} 020 (2009); P. Ho$\check{\text r}$ava, \textit{Phys. Rev. D} \textbf{79} 084008 (2009).\\\

[24] E. M. Lifshitz,  \textit{Zh. Eksp. Teor. Fiz.} \textbf{11} 255 (1949). \\\

[25] C. Appignani et al, \textit{arXiv:} \textbf{0907.3121v2} [hep-th] (2009).\\\

[26] E. Kiritsis, \textit{Phys. Rev. D} \textbf{81} 044009 (2010).\\\

[27] M. Wang, \textit{Phys. Rev. D} \textbf{81} 083006 (2010).\\\

[28] J. Z. Tang, \textit{Phys. Rev. D} \textbf{81} 043515 (2010).\\\

[29] B. Chen et al, \textit{arXiv:} \textbf{0910.0338v1} [hep-th] (2009).\\\

[30] J. J. Penga, \textit{Eur. Phys. J. C} \textbf{66} 325 (2010).\\\

[31] M. Jamil et al, \textit{JCAP} \textbf{1007} 028 (2010).\\\

[32] M. R. Setare, \textit{arXiv:} \textbf{0909.0456v2} [hep-th] (2009).\\\

[33] A. Ali, \textit{arXiv:} \textbf{1004.2474v1} [astro-ph.CO] (2010).\\\

[34] Kh. Saaidi, \textit{Astrophys. Space Sci.} \textbf{332}  503 (2011).\\\

[35] M. R. Setare et al, \textit{arXiv:} \textbf{1003.0376v1}
[hep-th] (2010); \textit{JCAP} \textbf{02} 010  (2010).\\\

[36] M. Wang, \textit{Phys. Rev. D} \textbf{81} 083006 (2010).\\\

[37] E. Kiritsis, \textit{Phys. Rev. D} \textbf{81} 044009 (2010).\\\

[38] S. Mukohyama, \textit{arXiv:} \textbf{0905.3563v4} [hep-th] (2009).\\\

[39] E. N. Saridakis, \textit{Eur. Phys. J. C}
\textbf{67}  229  (2010); E. N. Saridakis, \textit{ Eur. Phys. J. C} \textbf{67} 229 (2010).\\\

[40] T. Nishioka, \textit{Class. Quant. Grav.} \textbf{26}  242001 (2009).\\\

[41] Z. G. Huang et al, \textit{Class. Quant. Grav.} \textbf{23} 6215 (2006).\\\

[42] L. Amendola, \textit{Phys. Rev. D} \textbf{62} 043511 (2000).\\\

[43] H. Q. Lu et al, \textit{arXiv:} \textbf{0409309} [hep-th] (2004).\\\

[44] Z. G. Huang, \textit{ Astrophys. Space Sci.} \textbf{315} 175
(2008).\\\

[45] Z. G. Huang, \textit{ Astrophys. Space Sci.} \textbf{ 331} 331 (2011).\\\

[46] L. Amendola et al,  \textit{Phys. Rev. D} \textbf{77} 123526  (2008).\\\

[47] I. P. Neupane and H. Trowland, \textit{Int. J. Mod. Phys. D} \textbf{19} 367 (2010).\\\

[48] L. Nan et al, \textit{ Chin. Phys. Lett.} \textbf{26} 069501 (2009).\\\

[49] F. Piazza and S. Tsujikawa, \textit{JCAP} \textbf{004} 0407 (2004).\\\

[50] R. Dick, \textit{arXiv:} \textbf{9609190v2} [hep-th] (1996).\\\

[51] Y. M. Cho and Y. Y. Keum, \textit{Mod. Phys. Lett. A} \textbf{13} 109 (1998).\\\

[52] Y. M. Cho. and J. H. Kim, \textit{Phys. Rev. D} \textbf{79} 023504 (2007).\\\

[53] M. Susperregi, \textit{Phys. Rev. D} \textbf{68} 103509 (2003).\\\

[54] T. Damour et al, \textit{Phys. Rev. Lett.} \textbf{64} 2 (1990).\\\

[55] T. Biswas et al, \textit{Phys. Rev. D} \textbf{74} 063501 (2006).\\\

[56] G. Plcclnelli et al, \textit{Phys. Lett. B} \textbf{277} 58 (1992).\\\

\end{document}